\documentstyle[amssymb,aps,12pt,epsfig]{revtex}

\begin{document}
\author{H. L. Haroutyunyan and G. Nienhuis}
\address{Huygens Laborotarium, Universiteit Leiden, Postbus 9504, \\
2300 RA Leiden, The Netherlands}
\title{Resonances for coupled Bose-Einstein Condensates}
\maketitle

\begin{abstract}
We study some effects arising from  periodic modulation of the
asymmetry and the barrier height of a two-well potential
containing a Bose-Einstein condensate. At certain modulation
frequencies the system exhibits resonances, which may lead to
enhancement of the tunneling rate between the wells and which can
be used to control the particle distribution among the wells. Some
of the effects predicted for a two-well system can be carried over
to the case of a Bose-Einstein condensate in an optical lattice.
\end{abstract}

\section{Introduction}

Since the experimental realization of Bose-Einstein condensates
(BEC) one has considered the possibility of extending one-mode
models to two or more modes \cite{Andrews}. This raises the issue
of the relative phase between modes. As indicated by several
authors \cite{Sachdev}, a two-well BEC may exhibit features that
are not covered by the semiclassical description in terms of the
Gross-Pitaevski equation. These features are significant at low
particle numbers and for strong interactions. In previous work, we
discussed some aspects of the dynamics of a two-well BEC in the
strong-interaction regime \cite{analogy}. This is close to the
experimental situation for a BEC in a double-well trap, designed
in Ref. \cite{Tiecke}.

A sensitive way to probe the properties of a BEC in a double-well
potential with strong interatomic interactions is to look for
resonant behavior when a parameter of the system is periodically
modulated. The response of the system may be expected to be very
sensitive to the value of the modulation frequency in the
neighborhood of a resonance. A periodic perturbation can be
implemented in various ways. One example would be periodically
modulating the trapping potential. Salmond et al \cite{Salmond}
study a numerical model of a double-well potential with
periodically modulated coupling between the wells. This
semiclassical analysis reveals the existence of uncoupled regions
with chaotic and regular motion. The inclusion of the quantum
nature of the evolution leads to transitions between these
regions. Another type of periodic perturbation can be imposed by
periodically modulating the energy difference between the ground
states in the two wells.

Periodic modulations are known to give rise to dynamical
localization in some cases. This effect has been widely discussed
in the literature in the case a particle in a periodic potential,
such as an electron in a crystal or an atom in an optical lattice
\cite{Kenkre} and \cite{blochoscil}. When the particle also feels
a uniform force in addition to the lattice potential (a tilted
optical lattice), it is found to have an infinite discrete set of
equidistant energy levels, with a level separation that is
determined by the strength of the uniform force \cite{blochoscil}.
A variation of the magnitude of the uniform force affects the
phase of the state. So, when this magnitude is periodically
modulated, resonances may be expected. The population in one well
is described by adding the amplitudes for arriving at this well
from various other wells, each one with a different phase shift.
In the long time limit, when the time of observation is long
compared to the period of the modulation, this gives rise to
destructive interference, leading to a suppression of the net
tunneling rate. Hence, the asymptotic distribution over the wells
coincides with the initial one, and dynamical localization has
been realized.

Another example of dynamical localization arises for a single
two-level atom in a driving field with a periodically modulated
detuning \cite{Agarwal}. For certain ratios of the modulation
frequency and the strength of the field, the atom is localized in
its ground state. The time scale is restricted by the modulation
frequency.

In the present case of a BEC in a two-well potential with a fixed
total number of atoms, the state space is finite dimensional. In
the SU(2) representation of the operator algebra, the Hamiltonian
has a quadratic term due to the interatomic interactions. So, in
this sense the system is quite different from that of an atom in a
tilted lattice, with its infinite number of states and a
Hamiltonian that is linear in the SU(2) operators. Still, there
are some obvious similarities: the discrete structure of the
energy and the presence of interatomic interactions and tunneling
between wells as competing processes. Therefore, we expect
interesting effects also in the two-well case when the energy
difference or the hopping between wells is periodically modulated.
From a formal point of view, the analysis in the present paper may
be regarded as a generalization of the process of dynamical
localization for the Hamiltonian with a quadratic term.
Specifically, this paper considers the possibilites of coherent
control of a BEC in a double-well potential by using any kind of
time-periodic perturbation.

\section{BEC in a double potential well}

We describe a BEC in a double potential well in terms of a
one-particle
Hamiltonian $H^{(1)}$ and a two-particle interaction $U({\vec{r}},{\vec{r}}%
^{\prime })$. The states $\left| 1\right\rangle $ and $\left|
2\right\rangle
$ are the the localized ground states in either well, with wave functions $%
\psi _{1}({\vec{r}})$ and $\psi _{2}({\vec{r}})$. On the basis of
the states $\left| 1\right\rangle $ and $\left| 2\right\rangle $,
the one-particle Hamiltonian has the matrix elements
\begin{equation}
\left\langle 1\right| H^{(1)}\left| 1\right\rangle =-\left\langle
2\right| H^{(1)}\left| 2\right\rangle =\hbar \varepsilon
/2,\left\langle 1\right| H^{(1)}\left| 2\right\rangle
=\left\langle 2\right| H^{(1)}\left| 1\right\rangle =-\hbar \delta
/2\;.  \label{couplingconst}
\end{equation}
In the case that $\varepsilon =0$, the coupling between the wells
lifts their degeneracy, and creates an energy splitting $\hbar
\delta $ between
the even ground state $\left| g\right\rangle $ and the odd excited state $%
\left| e\right\rangle $, defined by
\begin{equation}
\left| g\right\rangle =\frac{1}{\sqrt{2}}(\left| 1\right\rangle
+\left| 2\right\rangle );\;\left| e\right\rangle
=\frac{1}{\sqrt{2}}(\left| 1\right\rangle -\left| 2\right\rangle
)\;.  \label{rbec1}
\end{equation}
When we restrict ourselves to these two states, the field operator
in second quantization has the standard form

\begin{equation}
\widehat{\Psi }(\overrightarrow{r})=\widehat{a}_{1}\psi _{1}({\vec{r}})+%
\widehat{a}_{2}\psi _{2}({\vec{r}}),  \label{psi}
\end{equation}
with $\widehat{a}_{i}$ the one-particle annihilation operator in
the two states, which together with the corresponding creation
operators obey the bosonic commutation rules. When we substitute
this expression in the formal expression
\begin{equation}
\widehat{H}=\int d\overrightarrow{r}\widehat{\Psi }^{\dagger }(%
\overrightarrow{r})H^{(1)}\widehat{\Psi }(\overrightarrow{r})+\frac{1}{2}%
\int d\overrightarrow{r}d\overrightarrow{r}^{\prime }\widehat{\Psi }%
^{\dagger }(\overrightarrow{r})\widehat{\Psi }^{\dagger }(\overrightarrow{r}%
^{\prime })U({\vec{r}},{\vec{r}}^{\prime })\widehat{\Psi }(\overrightarrow{r}%
)\widehat{\Psi }(\overrightarrow{r}^{\prime }).  \label{secquant}
\end{equation}
for the second-quantized Hamiltonian, we find

\begin{equation}
\widehat{H}=\sum_{i,k}\hbar \left\langle i\right| H^{(1)}\left|
k\right\rangle \widehat{a}_{i}^{\dagger }\widehat{a}_{k}+\frac{1}{2}%
\sum_{i,k,l,m}\hbar \left\langle i,k\right| U\left|
l,m\right\rangle \kappa
_{i,k,l,m}\widehat{a}_{i}^{\dagger }\widehat{a}_{k}^{\dagger }\widehat{a}_{l}%
\widehat{a}_{m},  \label{rHamfield}
\end{equation}
where the indices $i$, $j$, $k$, $l=1$ or $2$, and the matrix
elements are taken between the states $\psi _{1}$ and $\psi _{2}$.
$.$

At sufficiently low energy, the two particle interaction is well
approximated by the contact potential
$U({\vec{r}},{\vec{r}}^{\prime })=(4\pi \hbar ^{2}a/m)\delta
({\vec{r}}-{\vec{r}}^{\prime })$, with $a$ the scattering length.
The function $\psi _{1}$ and $\psi _{2}$ states have the same
form, and we assume that they do not overlap. So we obtain the
following expression for the Hamiltonian
\begin{equation}
\widehat{H}=\frac{\hbar \varepsilon }{2}\left( \widehat{a}_{1}^{\dagger }%
\widehat{a}_{1}-\widehat{a}_{2}^{\dagger }\widehat{a}_{2}\right) -\frac{%
\hbar \delta }{2}\left( \widehat{a}_{1}^{\dagger }\widehat{a}_{2}+\widehat{a}%
_{2}^{\dagger }\widehat{a}_{1}\right) +\frac{\hbar \kappa
}{2}\left(
\widehat{a}_{1}^{\dagger }\widehat{a}_{1}^{\dagger }\widehat{a}_{1}\widehat{a%
}_{1}+\widehat{a}_{2}^{\dagger }\widehat{a}_{2}^{\dagger }\widehat{a}_{2}%
\widehat{a}_{2}\right) .  \label{rbec3}
\end{equation}
where the parameter $\kappa $ defined by
\begin{equation}
\hbar \kappa ={\frac{4\pi \hbar ^{2}a}{m}}\int d{\vec{r}}\;|\psi _{1}({\vec{r%
}})|^{4}\;  \label{rkappa}
\end{equation}
measures the strength of the interatomic interaction.

For convenience we express the Hamiltonian (\ref{rbec3}) in terms
of SU(2) operators by applying the standard Schwinger
representation of two modes. This leads to the definition
\begin{equation}
\widehat{J}_{z}={\frac{1}{2}}\left( \widehat{a}_{1}^{\dagger }\widehat{a}%
_{1}-\widehat{a}_{2}^{\dagger }\widehat{a}_{2}\right) \;,\;\widehat{J}_{+}=%
\widehat{a}_{1}^{\dagger }\widehat{a}_{2}\;,\;\widehat{J}_{-}=\widehat{a}%
_{2}^{\dagger }\widehat{a}_{1}.  \label{rbec5}
\end{equation}
These operators are related to the Cartesian components of a
fictitious
angular momentum by the standard relations $\widehat{J}_{\pm }=\widehat{J}%
_{x}\pm i\widehat{J}_{y}$. They obey the commutation rules for
angular momentum operators
\begin{equation}
\lbrack \widehat{J}_{z},\widehat{J}_{\pm }]=\pm \widehat{J}_{\pm }\;,\;[%
\widehat{J}_{+},\widehat{J}_{-}]=2\widehat{J}_{z}\;, \label{rbec6}
\end{equation}
which generate the SU(2) algebra. These operators commute with the
operator
for the total number of particles $\widehat{N}=\widehat{a}_{1}^{\dagger }%
\widehat{a}_{1}+\widehat{a}_{2}^{\dagger }\widehat{a}_{2}$. The Hamiltonian (%
\ref{rbec3}) can be rewritten in the form
\begin{equation}
\widehat{H}=\widehat{H}_{N}+\frac{\hbar \kappa }{4}\left( \widehat{N}^{2}-2%
\widehat{N}\right) \;,  \label{rHamop}
\end{equation}
where the $N$-particle Hamiltonian $\widehat{H}_{N}$ is defined by
\begin{equation}
\widehat{H}_{N}=\hbar \varepsilon \widehat{J}_{z}-\hbar \delta \widehat{J}%
_{x}+\hbar \kappa \widehat{J}_{z}^{2}.  \label{rHN}
\end{equation}
For a given number of particles $N$, the last term in Eq.
(\ref{rHamop}) is a
constant, and it suffices to consider the dynamics of the subspace of the $%
N+1$ number states $\left| n,N-n\right\rangle $, with $n=0,1,\dots N$, with $%
n$ particles in well $1$, and $N-n$ particles in well $2$. This
subspace has the structure of the angular momentum states, with
$J=N/2$, and the $2J+1$ magnetic quantum numbers $\mu =n-N/2$,
with $\mu =-J,-J+1,\dots ,J$. Note that $\mu $ is half the
difference of the particle number in two wells. For a given
particle number $N$ we represent the number states by the quantum
number $\mu $, so that $\left| \mu \right\rangle \equiv \left|
n,N-n\right\rangle $. The action of the operators $\widehat{J}_{0}$ and $%
\widehat{J}_{\pm }$ on the Fock states has the well-known behavior
\begin{equation}
\widehat{J}_{z}\left| \mu \right\rangle =\mu \left| \mu \right\rangle \;,\;%
\widehat{J}_{\pm }\left| \mu \right\rangle =\sqrt{\left( J\mp \mu
\right) \left( J\pm \mu +1\right) }\left| \mu \pm 1\right\rangle .
\label{rJpm}
\end{equation}
This also determines the action of the Cartesian operators
$\widehat{J}_{x}$ and $\widehat{J}_{y}$.

\section{Quantum states in two wells}

The Schwinger representation of the operators occurring in the
Hamiltonian suggests in a natural way various possible choices of
states of $N$ atoms in the two wells. Arecchi et al \cite{Arecchi}
introduced the spin coherent states (SCS) \cite{Perelomov}, in
analogy to the Glauber coherent state of a mode of the quantum
radiation field. The SCS follow from applying an arbitrary
rotation to the state $\left| \mu \right\rangle $ with $\mu =J$.
As rotation operator we take
\begin{equation}
\widehat{R}(\theta ,\phi )=\exp (-i\phi \widehat{J}_{z})\exp
(-i\theta
\widehat{J}_{y})\exp (i\phi \widehat{J}_{z})=\exp [-i\theta (\widehat{J}%
_{y}\cos \phi -\widehat{J}_{x}\sin \phi )],  \label{rrotation}
\end{equation}
which represents a rotation over an angle $\theta $, around an axis in the $%
xy$-plane, specified by the angle $\phi $ with the $y$-axis. The
SCS $\left| \theta ,\phi ;J\right\rangle $ is
\begin{equation}
\left| \theta ,\phi ;J\right\rangle =\widehat{R}(\theta ,\phi
)\left| J\right\rangle  \label{Disp1}
\end{equation}
which is also the eigenstate with eigenvalue $J$ of the component $%
\overrightarrow{u}\cdot \widehat{\overrightarrow{J}}$ of the
angular-momentum vector in the direction $\overrightarrow{u}$
specified by the polar angle $\theta $ and the azimuthal angle
$\phi $. Just as the Glauber coherent states of a mode with
annihilation operator $\widehat{c}$ can be obtained by acting with
a displacement operator $\widehat{D}\left(
\zeta \right) =$ $\exp \left( \zeta \widehat{c}^{\dagger }-\zeta ^{*}%
\widehat{c}\right) $ on the vacuum state, the SCS follows by a rotation $%
\widehat{R}(\theta ,\phi )=\exp \left( \zeta \widehat{J}_{-}-\zeta ^{*}%
\widehat{J}_{+}\right) $ with $\zeta =\left( \theta /2\right) \exp
\left( i\phi \right) $, acting on the state $\left| J\right\rangle
$. When we view
this state $\left| J\right\rangle $ as the ground state, the operator $%
\widehat{J}_{+}$ is analogous to the annihilation operator, since $\widehat{J%
}_{+}\left| J\right\rangle =0$. An essential difference between
the two cases is, of course, that the state space of a radiation
mode has infinite
dimensions, while the dimension of the angular-momentum state space is $2J+1$%
.

In our case, the analogy is carried one step further, since the
SCS defined by (\ref{Disp1}) do not represent angular-momentum
states, but refer to the
states of $N$ atoms, distributed over two potential wells. The ground state $%
\left| J\right\rangle $ represents the state with all particles in
the first well. When we substitute the identity $\left|
J\right\rangle =\left( \widehat{a}_{1}^{\dagger }\right)
^{N}\left| vac\right\rangle /\sqrt{N!}$ with $N=2J$ into the
right-hand side of (\ref{Disp1}), we obtain an expression for the
SCS in the language of the two wells, in the form
\begin{equation}
\left| \theta ,\phi ;J\right\rangle =\frac{1}{\sqrt{N!}}\left( \cos \frac{%
\theta }{2}\text{ }\widehat{a}_{1}^{\dagger }+e^{i\phi }\sin \frac{\theta }{2%
}\text{ }\widehat{a}_{2}^{\dagger }\right) ^{N}\left|
vac\right\rangle . \label{Disp2}
\end{equation}
We can interpret (\ref{Disp2}) as a state with $N$ atoms in the
one-particle superposition state
\[
\cos \frac{\theta }{2}\text{ }\left| 1\right\rangle +e^{i\phi }\sin \frac{%
\theta }{2}\text{ }\left| 2\right\rangle
\]
of the two wells. A special case arises for $\theta =\pi /2$, when
the average populations of the two wells are the same. Then the
state (\ref {Disp2}) describes a collective mode from two
interfering sources of equal intensity, and its expansion in
number states is given by
\begin{equation}
\left| \pi /2,\phi ;J\right\rangle
=\frac{1}{2^{N/2}}\sum_{n=0}^{N}\left(
\begin{array}{l}
N \\
n
\end{array}
\right) ^{1/2}e^{i\left( N-n\right) \phi }\left|
n,N-n\right\rangle \label{rphase}
\end{equation}
Such a state can be considered as a state with a well-defined
phase difference $\phi $. The atom distribution over the two wells
is binomial, and they have been termed phase states (PS) of a
two-mode boson system in Ref. \cite{Castin}. For simplicity, we
suppress the value $\pi /2$ in this case, and we simply denote the
PS as $\left| \phi ;J\right\rangle $ . Upon rotation around the
$z$-axis, a PS\ transforms as
\begin{equation}
\exp (-i\alpha \widehat{J}_{z})\left| \phi ;J\right\rangle =\exp
(-i\alpha J)\left| \phi +\alpha ;J\right\rangle  \label{shift}
\end{equation}

The concept of Glauber coherent states of a radiation mode has
been generalized by de Oliveira et al \cite{de Oliveira}, who
introduced so called displaced coherent states defining them as a
displaced number state, rather than a displaced vacuum state. The
corresponding generalization of a SCS is found when the rotation
operator (\ref{rrotation}) acts on a number state $\left| \mu
\right\rangle $. The resulting displaced spin coherent states
(DSCS) are

\begin{equation}
\left| \theta ,\phi ;\mu \right\rangle =\widehat{R}(\theta ,\phi
)\left| \mu \right\rangle .  \label{Disp4}
\end{equation}
They are the eigenstates of the angular-momentum component $\overrightarrow{u%
}\cdot \widehat{\overrightarrow{J}}$ with eigenvalue $\mu $. In
the special case that $\theta =\pi /2$ and $\phi =0$, we find that
$\widehat{R}(\theta
,\phi )=\exp (-i\pi \widehat{J}_{y}/2)$, and this component is simply $%
\widehat{J}_{x}$. Its eigenstates are denoted as
\[
\left| \pi /2,0;\mu \right\rangle =\left| \mu \right\rangle _{x}.
\]
and they obey the eigenvalue relation $\widehat{J}_{x}\left| \mu
\right\rangle _{x}=\mu \left| \mu \right\rangle _{x}$. The state
$\left| \mu
\right\rangle _{x}$ describes a state with $J+\mu $ atoms in the even state $%
\left| g\right\rangle $, and $J-\mu $ atoms in the odd state
$\left| e\right\rangle $. These states are coupled by the ladder
operators
\begin{equation}
\widehat{J}_{x}^{\pm }\equiv \exp (-i\pi
\widehat{J}_{y}/2)\widehat{J}_{\pm }\exp (i\pi
\widehat{J}_{y}/2)=-\widehat{J}_{z}\pm i\widehat{J}_{y},
\label{Jxplus}
\end{equation}
according to the relations
\[
\widehat{J}_{x}^{\pm }\left| \mu \right\rangle _{x}=\sqrt{\left(
J\mp \mu \right) \left( J\pm \mu +1\right) }\left| \mu \pm
1\right\rangle _{x}.
\]

When $\theta =\pi /2$ and $\phi =\pi /2$, the DSCS are indicated
as
\[
\left| \pi /2,\pi /2;\mu \right\rangle =\left| \mu \right\rangle
_{y},
\]
which are eigenstates of $\overrightarrow{u}\cdot \widehat{\overrightarrow{J}%
}=\widehat{J}_{y}$, as specified by the relation
$\widehat{J}_{y}\left| \mu \right\rangle _{y}=\mu \left| \mu
\right\rangle _{y}$. The corresponding ladder operators are
\begin{equation}
\widehat{J}_{y}^{\pm }\equiv \exp (i\pi
\widehat{J}_{x}/2)\widehat{J}_{\pm }\exp (-i\pi
\widehat{J}_{x}/2)=\widehat{J}_{x}\mp i\widehat{J}_{z}.
\label{Jyplus}
\end{equation}
In the special case that $\mu =J$, the DSCS $\left| \mu
\right\rangle _{x}$ and $\left| \mu \right\rangle _{y}$ become the
SCS $\left| J\right\rangle _{x}$ and $\left| J\right\rangle _{y}$,
which are also the PS with $\phi =0$ and $\phi =\pi /2$,
respectively.

\section{Evolution in limiting cases}

For a given number $N$ of atoms, the evolution is characterized by
an evolution operator that is governed by the Hamiltonian
(\ref{rHN}), and that obeys the Schr\"{o}dinger equation
\begin{equation}
i\hbar \frac{d\widehat{U}}{dt}=\widehat{H}_{N}\widehat{U}.
\label{int1}
\end{equation}
In order to get an intuitive insight into the evolution, we first
consider two extreme cases, which are simple to understand. We
assume that the two wells have equal energy, so that $\varepsilon
=0$. If the interatomic interactions are negligible, the quadratic
term in (\ref{rHN}) can be skipped. For a possibly time-dependent
coupling strength $\delta $, the evolution operator is

\begin{equation}
\widehat{U}(t)=\exp \left( i\eta \left( t\right)
\widehat{J}_{x}\right) , \label{int15}
\end{equation}
with $\eta \left( t\right) =\int_{0}^{t}\delta (t^{^{\prime
}})dt^{^{\prime
}}$ the area of the coupling pulse. In the language of angular momentum, $%
\widehat{U}$ represent a rotation over an angle $-\eta $ around the $x$%
-axis. In this case, the states $\left| \mu \right\rangle _{x}$
are eigenstates of the evolution operator, so that these states
acquire only a phase factor $\exp \left( i\eta \left( t\right) \mu
\right) $. An initial state in the form of a single number state
$\left| \mu \right\rangle $ state gets rotated by the operator
(\ref{int15}) and evolves into a superpostion of number states. At
the instant that $\eta \left( t\right) =\pi /2$ an
initial number state has evolved into an eigenstate of the operator $%
\widehat{J}_{y}.$

Conversely, when the interatomic interactions are strong enough on
the scale of tunneling, the hopping between the wells can get
suppressed \cite{analogy}. Now, a single number state $\left| \mu
\right\rangle $ only acquires a phase factor $\exp (-i\kappa \mu
^{2}t)$. The evolution operator takes the form
\begin{equation}
\widehat{U}\left( t\right) =e^{-i\kappa \widehat{J}_{z}^{\text{
}2}t}, \label{int16}
\end{equation}
which cannot be conceived as a rotation in the angular-momentum
space. Since the eigenvalues of $\widehat{J}_{z}^{2}$ are
discrete, the evolution (\ref {int16}) has revivals. First we
consider the situation that the number of
particles $N$ is even, so that the eigenvalues $\mu $ of $\widehat{J}_{z}^{%
\text{ }}$ are integer. Then the eigenvalues of $\widehat{U}$ are
$\exp (-i\kappa \mu ^{2}t)=1$ when $t=mT$, with $m$ an integer,
and $T=2\pi /\kappa $ . At these times the initial state is
reproduced, which proves that the evolution of any initial state
is time periodic, with period $T$. For a time $t=T/2$, which is
half the period, the eigenvalues of $\widehat{U}(T/2)$ are $\exp
(-i\pi \mu ^{2})=(-1)^{\mu }$, which proves that the evolution
operator at this instant is equal to
\[
\widehat{U}(T/2)=\exp (-i\pi \widehat{J}_{z}).
\]
For an initial PS $\left| \Psi (0)\right\rangle =\left| \phi
;J\right\rangle $, we find that the state at the time $t=T/2$ is
\[
\left| \Psi (T/2)\right\rangle =\exp (-i\pi J)\left| \phi
;J\right\rangle ,
\]
which is just the opposite PS. At other instants of time, that are
a simple rational fraction of $T$, an initial PS can be
transformed into a linear
combination of a few PS. For $t=T/4$, the relevant eigenvalues of $\widehat{U%
}$ can be rewritten as

\[
\exp (-i\pi \mu ^{2}/2)=\frac{1}{\sqrt{2}}\left( e^{-i\pi /4}+\exp
(-i\pi \widehat{J}_{z})e^{i\pi /4}\right) .
\]
The corresponding expression for the evolution operator is then
\[
\widehat{U}(T/4)=\frac{1}{\sqrt{2}}\left[ e^{-i\pi /4}+e^{i\pi
/4}\exp (-i\pi \widehat{J}_{z})\right] .
\]
For the same initial state $\left| \Psi (0)\right\rangle =\left|
\phi ;J\right\rangle $, we apply Eq. (\ref{shift}), and arrive at
the result for
the state at $t=T/4$%
\begin{equation}
\left| \Psi (T/4)\right\rangle =\frac{1}{\sqrt{2}}\left[ e^{-i\pi
/4}\left| \phi ;J\right\rangle +e^{i\pi /4}e^{-i\pi J}\left| \phi
+\pi ;J\right\rangle \right] ,  \label{int17}
\end{equation}
which is the linear superposition of two PS's. For times $t$ that
are equal to other simple rational fractions of the period $T$
($t=T/3$, $T/5$,..) a superposition of more PS's is found. One may
use the fact that the eigenvalues $\exp (-i\kappa \mu ^{\text{
}2}t)$ of $\widehat{U}$ are periodic in $\mu $ with some integer
period $p$. Therefore these eigenvalues can be expressed as a
finite Fourier series in powers of $\exp (2\pi i\mu
/p) $, which is equivalent to expressing the evolution operator $\widehat{U}%
(t)$ as a finite sum of rotations around the $z$-axis.

When the number $N$ of particles is odd, so that the values of $J$
and $\mu $ are half-integer, full revival of the initial state is
again found after one period $t=T$. In fact, since $2\mu $ is an
odd number, $(4\mu ^{2}-1)/4$ is always an even integer, and it
follows that both at time $T$ and $T/2$, the evolution operator is
just a phase factor
\[
\widehat{U}(T)=\exp (-i\pi /2),\widehat{U}(T/2)=\exp (-i\pi /4).
\]
Hence, apart from a phase factor, full revival is found already at
half the time $T$. In order to obtain the evolution operator at
the time $t=T/4$, it is convenient to use the identity for half
integer values of $\mu $
\[
\exp (-i\pi \mu ^{2}/2)=\frac{1}{\sqrt{2}}e^{-i\pi /8}\left(
e^{i\pi \mu /2}+e^{-i\pi \mu /2}\right) .
\]
For the evolution operator this gives the expression
\[
\widehat{U}(T/4)=\frac{1}{\sqrt{2}}e^{-i\pi /8}\left[ \exp (i\pi \widehat{J}%
_{z}/2)+\exp (-i\pi \widehat{J}_{z}/2)\right] .
\]
For the initial PS $\left| \Psi (0)\right\rangle =\left| \phi
;J\right\rangle $, we obtain for the state vector at time $T/4$%
\[
\left| \Psi (T/4)\right\rangle =\frac{1}{\sqrt{2}}e^{-i\pi
/8}\left[ e^{i\pi J/2}\left| \phi -\pi /2;J\right\rangle +e^{-i\pi
J/2}\left| \phi +\pi /2;J\right\rangle \right] .
\]
Revivals of the state of a BEC have been observed in an optical
lattice\cite {Greiner2}.

\section{Periodic modulation of energy difference}

A simple example of a periodic modulation of the two-well system
is to include a time-varying energy difference between the two
wells. This is realized by substituting in the $N$-particle
Hamiltonian (\ref{rHN}) the harmonically varying parameter
$\varepsilon \left( t\right) =\varepsilon _{1}\cos \omega t$,
while $\delta $ and $\kappa $ remain constant. It is convenient to
describe the evolution in an interaction picture that removes the
diagonal terms in the Hamiltonian. We introduce the transformed
state vector $\left| \Psi ^{\prime }(t)\right\rangle $ by the
relation
\begin{equation}
\left| \Psi \left( t\right) \right\rangle =\widehat{T}(t)\left|
\Psi ^{^{\prime }}\left( t\right) \right\rangle ,  \label{rtrans}
\end{equation}
where the state vector $\left| \Psi \left( t\right) \right\rangle
$ obeys the Schr\"{o}dinger equation with the Hamiltonian
(\ref{rHN}), and the transformation operator $\widehat{T}(t)$ is
defined by
\begin{equation}
\widehat{T}(t)=\exp \left[ -i\theta \left( t\right) \widehat{J}_{z}-i\kappa t%
\widehat{J}_{z}^{2}\right] ,  \label{Tt}
\end{equation}
with $\theta (t)=\int_{0}^{t}dt^{^{\prime }}\varepsilon \left(
t^{^{\prime }}\right) =\varepsilon _{1}\left( \sin \omega t\right)
/\omega $. Notice that the transformed state $\left| \Psi ^{\prime
}(t)\right\rangle $ has the same distribution over the number
states $\left| \mu \right\rangle $ as the actual state $\left|
\Psi (t)\right\rangle $. The transformed Schr\"{o}dinger equation
has the standard form
\begin{equation}
i\hbar \frac{d\left| \Psi ^{^{\prime }}\left( t\right) \right\rangle }{dt}=%
\widehat{H}^{\prime }(t)\left| \Psi ^{^{\prime }}\left( t\right)
\right\rangle .  \label{intpicture}
\end{equation}
An explicit form of the transformed Hamiltonian
\begin{equation}
\widehat{H}^{\prime }(t)=-\hbar \delta \widehat{T}^{\dagger }(t)\widehat{J}%
_{x}\widehat{T}(t)  \label{Hprime}
\end{equation}
follows from the general transformation rule \cite{Kitagawa}
\begin{equation}
f(\widehat{J}_{z})\widehat{J}_{+}=\widehat{J}_{+}f(\widehat{J}_{z}+1).
\label{LinearTransform}
\end{equation}
Ths relation (\ref{LinearTransform}) holds for any analytical
function $f$
of the operator $\widehat{J}_{z}$. After substituting (\ref{LinearTransform}%
) into (\ref{Hprime}), we arrive at the result
\[
\widehat{H}^{\prime }(t)=-\frac{\hbar \delta }{2}\left[ \widehat{J}%
_{+}e^{i\theta (t)+i\kappa t\left( 2\widehat{J}_{z}+1\right) }+\text{H.c.}%
\right] .
\]
After a Fourier expansion of the exponentials, we find
\begin{equation}
\widehat{H}^{\prime }(t)=-\frac{\hbar \delta }{2}\sum_{n=-\infty
}^{\infty \text{ }}J_{n}\left(
{\displaystyle {\varepsilon _{1} / \omega }}%
\right) \left( \widehat{J}_{+}e^{it\left[ \kappa \left( 2\widehat{J}%
_{z}+1\right) +n\omega \right] }+\text{H.c.}\right) .
\label{Hresonant}
\end{equation}
The form (\ref{Hresonant}) of the operator $\widehat{H}^{\prime
}(t)$ allows
a clear physical interpretation. The oscillating energy difference $%
\varepsilon (t)$ is equivalent to a series of harmonic couplings
between the wells with equally spaced driving frequencies $n\omega
$. The amplitude for each harmonic is proportional to the Bessel
function of the corresponding order. So the effective coupling
between the number states $\left| \mu \right\rangle $ depends
strongly on the frequency.

The Hamiltonian $\widehat{H}^{\prime }(t)$ contains only
non-vanishing
elements coupling neighboring number states $\left| \mu \right\rangle $ and $%
\left| \mu +1\right\rangle $. A resonance occurs for the $n$th
harmonic when
\begin{equation}
n\omega +\kappa (2\mu +1)=0,  \label{rescond}
\end{equation}
which requires that $\kappa \left( 2\mu +1\right) /\omega $ is an
integer.

The strength of this coupling is $-\Omega _{\mu}J_{n}(\varepsilon
_{1}/e\omega )/2$, with
\begin{equation}
\Omega _{\mu }=\delta \sqrt{\left( J-\mu \right) \left( J+\mu
+1\right) }. \label{omega}
\end{equation}
The effective coupling by the $n$th harmonic is measured by the
parameter
\begin{equation}
U_{\mu }^{n}=\frac{\Omega _{\mu }}{n\omega +\kappa (2\mu
+1)}J_{n}\left(
{\displaystyle {\varepsilon _{1} \over \omega }}%
\right) ,  \label{effectivecoupling}
\end{equation}
which is the ratio of the coupling strength and the detuning from
resonance for the transition. Whenever $\left| U_{\mu }^{n}\right|
\ll 1$, the coupling is weak.

When the oscillation frequency $\omega $ is large compared with
the maximal diagonal frequency splitting $\kappa (2J+1)$, all the
time-dependent
couplings are weak, and the dominant coupling term is the static one with $%
n=0$. The effect of the modulated energy difference is then that
the coupling term is reduced by the factor $J_{0}(\varepsilon
_{1}/\omega )$. In the high-frequency limit $\omega \gg
\varepsilon _{1}$, this factor is one, and we recover the case of
a static and symmetric double-well potential with $\varepsilon
_{1}=0$.

A simple isolated resonance between two number states can occur
involving the states $\left| \mu \right\rangle $ with $\mu =-J$ or
$\mu =J$, since these can be coupled to only one other state.
Suppose that at $t=0$ all atoms are in one of the two wells, so
that
\begin{equation}
\left| \Psi \left( t=0\right) \right\rangle =\left|
-J\right\rangle . \label{initialcond}
\end{equation}

\begin{figure}[!ht]
\centerline{\psfig{figure=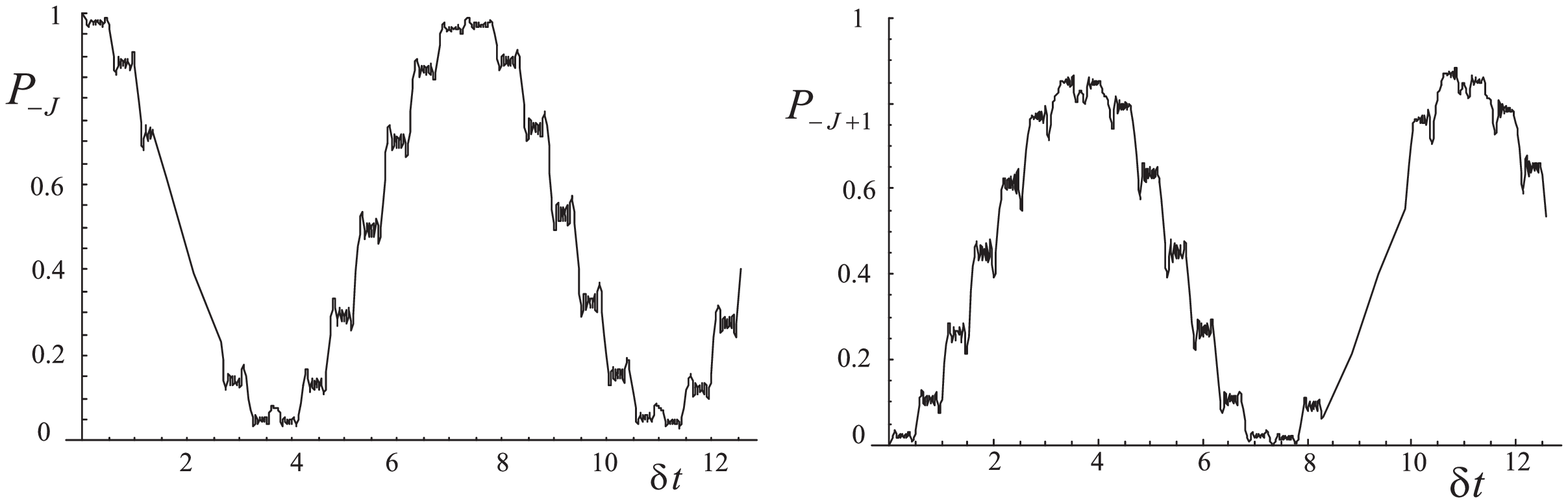,height=4cm}} \caption{Time
dependence of the populations $P_{\mu }$ for the state $\left| \mu
\right\rangle =\left| -J\right\rangle $ and $\left| \mu
\right\rangle =\left| -J+1\right\rangle .$ The parameters are
taken as $\delta /\kappa =0.25$, $\varepsilon _{1}/\kappa =14$,
$\omega /\kappa =3$, $N=2J=16$.} \label{resf1}
\end{figure}

If one chooses the frequency $\omega $ such that the resonance condition (%
\ref{rescond}) holds at certain integer $n_{0},$ the corresponding
harmonic can be made dominant. Indeed, provided that $\left|
U_{\mu }^{n}\right| \ll 1 $ for $\mu =-J+1$, for all $n$, coupling
to other states is weak, and we have an effective two-level
system. This is possible provided that at resonance $\omega $ is
large compared with $\kappa $, which in turn is large compared
with the coupling parameter $\delta $. This is demonstrated in
Fig. \ref{resf1}, where oscillations between the states $\left| -J\right\rangle $ and $%
\left| -J+1\right\rangle $ are displayed for the initial state
(\ref {initialcond})$.$ This means that a single atom out of $N$
atoms resonantly oscillates between the wells. Upon decreasing the
coupling between the wells, the rate of off-resonant coupling is
decreasing, so one approaches ideal Rabi oscillations between
resonant levels. Weaker coupling implies a larger oscillation
period. The two-level behavior can only occur for a system with a
nonlinear term $\widehat{J}_{z}^{2}$, since for a linear system
the various transitions are simultaneously in resonance \cite
{Kenkre} and \cite{blochoscil}.

In the case that the modulation frequency $\omega $ is of the same order as $%
\kappa $, resonances on the different transitions can coincide,
and the initial state (\ref{initialcond}) can spread out over many
number states. For example, in the simple case that $\omega
=\kappa $, the resonance condition (\ref{rescond}) shows that for
each value of $\mu $, there is a harmonic $n=-(2\mu +1)$ that is
resonant, and the population spreads out over all number states.

\begin{figure}[!ht]
\centerline{\psfig{figure=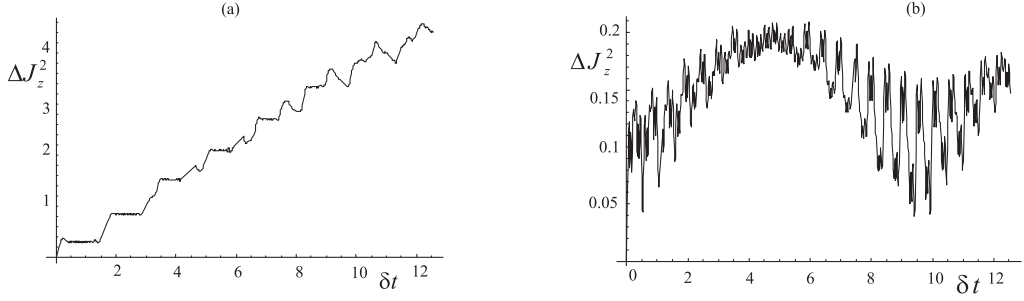,height=4cm}}
\caption{Time dependence of the fluctuation $\left\langle \widehat{J}%
_{z}\right\rangle ^{2}$ of operator $\widehat{J}_{z}$ at resonance (a) $%
\omega =\kappa $ and out of resonance (b) $\omega =6\kappa .$ The
other parameters are taken the same as in the previous figure.}
\label{resf2}
\end{figure}

The difference with the high-frequency case is demonstrated in
Fig. \ref{resf2}, where we plot the fluctuations $\Delta
J_{z}^{2}$ of $\widehat{J}_{z}$ as a function of time, for the
initial condition (\ref {initialcond}), for $\omega =4\kappa $ (a)
and $\omega =\kappa $ (b). In the first case, the fluctuations
remain limited. In the second case, a resonance occurs on each
transition, and $\Delta J_{z}^{2}$ continues to increase. Even for
a very small coupling between wells, resonances designed in such a
way can lead to enhancement in the tunneling rate. This is close
to the experimental situation for the double-well trap presented
in Ref. \cite {Tiecke}. Again, this situation is specific for a
system with a non-linear term $\widehat{J}_{z}^{2}$ in the
Hamiltonian, since for a linear system various transitions have
the same effective coupling. Since the coupling is proportional to
$J_{n}\left( \varepsilon _{1}/\omega \right) $, a resonant
transition can be turned off by setting the ratio $\varepsilon
_{1}/\omega $ equal to a zero of the Bessel function.

\begin{figure}[!ht]
\centerline{\psfig{figure=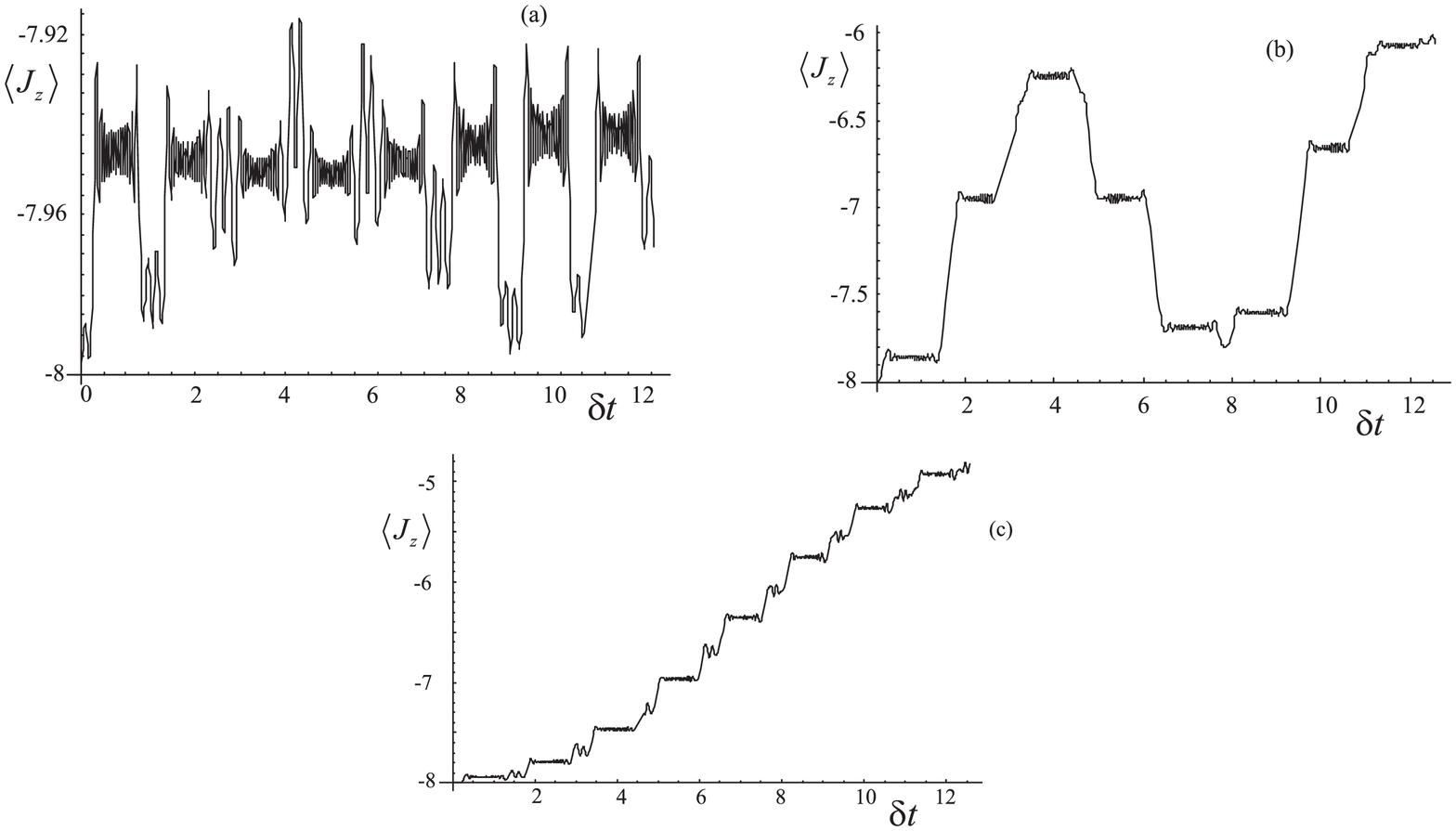,height=7cm}} \caption{Time
dependence of the expectation value of $\widehat{J}_{z}$ operator
is plotted at resonance $\omega =\kappa $. The ratio $\varepsilon
_{1}/\omega $ is chosen such that (a) $\varepsilon _{1}/\omega =24.26918$ ($%
J(_{13}\left( \varepsilon _{1}/\omega \right) =0$) ,  (b)
$\varepsilon _{1}/\omega =23.2759$ ($J_{11}\left( \varepsilon
_{1}/\omega \right) =0)$ (c) $\varepsilon _{1}/\omega =24.93493$
($J_{7}\left( \varepsilon _{1}/\omega \right) =0$)$.$ The total
number of particles is chosen $N=16$.} \label{resf3}
\end{figure}

This can be used to restrict the evolution to a limited number of
states, thereby locating a desired number of particles in one of
the wells. We demonstrate this idea in Fig. \ref{resf3}. There we
start from the same initial condition (\ref{initialcond}) and see
that $11
$ atoms out of $16$ are localized in the left well if one chooses the ratio $%
\varepsilon _{1}/\omega $ such that $J_{7}\left( \varepsilon
_{1}/\omega \right) =0$ (Fig. \ref{resf3}a). Then, taking
$J_{11}\left( \varepsilon _{1}/\omega \right) =0$, or
$J_{15}\left( \varepsilon _{1}/\omega \right) =0$, one can
localize $13$ or $16$ particles in one of the wells (Figs.
\ref{resf3}b and \ref{resf3}c).

\section{Generalization to an optical lattice}

The discussion of the previous section for two wells with an
energy difference can be generalized to the case of a multimode
system, consisting of a chain of potential wells. As a model, we
take a BEC in a tilted optical lattice \cite{blochoscil}. As
usual, we neglect the higher bands in the lattice, and we consider
only a BEC trapped in the lowest energy band, that roughly
speaking is composed of the ground states in all the wells \cite
{Jaksch}. If one takes the Wannier states $\left| l\right\rangle $ with $%
l=\ldots -2$, $-1$, $0$, $1$, $2$, $\ldots $as the basis of
one-particle states, the Hamiltonian in second quantization is a
direct generalization of Eq. (\ref{rbec3}) for two wells, and it
takes the form
\begin{equation}
\widehat{H}_{BH}=-\frac{\hbar \delta }{2}\sum_{l}\left( \widehat{a}%
_{l+1}^{\dagger }\widehat{a}_{l}+\widehat{a}_{l}^{\dagger }\widehat{a}%
_{l+1}\right) +\frac{\hbar \kappa }{2}\sum_{l}\widehat{a}_{l}^{\dagger }%
\widehat{a}_{l}^{\dagger }\widehat{a}_{l}\widehat{a}_{l}+\hbar
\varepsilon \left( t\right) \sum_{l}l\widehat{a}_{l}^{\dagger
}\widehat{a}_{l}, \label{BoseHubbard}
\end{equation}
where $\widehat{a}_{l\text{ }}\left( \widehat{a}_{l\text{
}}^{\dagger }\right) $ are bosonic annihilation (creation)
operators in a single Wannier state, $\delta $ and $\kappa $ are
the obvious multimode generalizations of two-mode definitions for
the nearest neighbor coupling and interaction constant
(\ref{couplingconst}, \ref{rkappa}), $\varepsilon $ is the energy
difference in frequency units between neighboring Wannier states,
which determines the uniform force. This Hamiltonian defines the
so-called Bose-Hubbard model.

The time evolution in a lattice is governed by the time-dependent
Schr\"{o}dinger equation
\begin{equation}
i\hbar \frac{d\left| \Psi \left( t\right) \right\rangle }{dt}=\widehat{H}%
_{BH}\left| \Psi \left( t\right) \right\rangle .
\label{EvolutionEq1}
\end{equation}
The uniform force and the interatomic interaction can be
eliminated by the substitution
\[
\left| \Psi _{BH}\left( t\right) \right\rangle
=\widehat{T}_{BH}(t)\left| \Psi _{BH}^{^{\prime }}\left( t\right)
\right\rangle ,
\]
with
\[
\widehat{T}_{BH}(t)=\exp \left( -i\theta \left( t\right) \sum_{l}l\widehat{a}%
_{l}^{\dagger }\widehat{a}_{l}-\frac{i\kappa t}{2}\sum_{l}\widehat{a}%
_{l}^{\dagger }\widehat{a}_{l}^{\dagger }\widehat{a}_{l}\widehat{a}%
_{l}\right)
\]
and $\theta \left( t\right) =\int_{0}^{t}dt^{^{\prime
}}\varepsilon \left( t^{^{\prime }}\right) $ is the area of pulse.
The Schr\"{o}dinger equation for the transformed state $\left|
\Psi _{BH}^{^{\prime }}\left( t\right) \right\rangle $ follows by
using the transformation properties of the annihilation operator
\[
\widehat{T}_{BH}^{\dagger }(t)\widehat{a}_{l}\widehat{T}_{BH}(t)=\widehat{a}%
_{l}\exp \left( -il\theta (t)-i\kappa t\left( \widehat{a}_{l}^{\dagger }%
\widehat{a}_{l}-1\right) \right) =\exp \left( -il\theta (t)-i\kappa t%
\widehat{a}_{l}^{\dagger }\widehat{a}_{l}\right) \widehat{a}_{l},
\]
which leads to the identity
\[
\widehat{T}_{BH}^{\dagger }(t)\widehat{a}_{l+1}^{\dagger }\widehat{a}_{l}%
\widehat{T}_{BH}(t)=\widehat{a}_{l+1}^{\dagger
}\widehat{a}_{l}\exp \left[
i\theta (t)+i\kappa t\left( \widehat{a}_{l+1}^{\dagger }\widehat{a}_{l+1}-%
\widehat{a}_{l}^{\dagger }\widehat{a}_{l}+1\right) \right] .
\]
We obtain the evolution equation
\begin{equation}
i\hbar \frac{d\left| \Psi _{BH}^{\prime }\left( t\right) \right\rangle }{dt}=%
\widehat{H}_{BH}^{^{\prime }}\left| \Psi _{BH}^{\prime }\left(
t\right) \right\rangle ,  \label{EvolutionEq2}
\end{equation}
with the effective Hamiltonian
\begin{equation}
\widehat{H}_{BH}^{^{\prime }}=-\frac{\hbar \delta
}{2}\sum_{l}\left( \widehat{a}_{l+1}^{\dagger }\widehat{a}_{l}\exp
\left[ i\theta (t)+i\kappa
t\left( \widehat{a}_{l+1}^{\dagger }\widehat{a}_{l+1}-\widehat{a}%
_{l}^{\dagger }\widehat{a}_{l}+1\right) \right]
+\text{H.c.}\right) . \label{BoseHubbard''}
\end{equation}
For the case of a periodically modulated uniform force, described by $%
\varepsilon \left( t\right) =\varepsilon _{1}\cos \omega t$, this
Hamiltonian can be put in the form
\begin{equation}
\widehat{H}_{BH}^{^{\prime }}=-\frac{\hbar \delta }{2}\sum_{l}\sum_{n}J_{n}%
\left( \frac{\varepsilon _{1}}{\omega }\right) \left( \widehat{a}%
_{l+1}^{\dagger }\widehat{a}_{l}\exp \left[ in\omega t+i\kappa
t\left(
\widehat{a}_{l+1}^{\dagger }\widehat{a}_{l+1}-\widehat{a}_{l}^{\dagger }%
\widehat{a}_{l}+1\right) \right] +\text{H.c.}\right) .
\label{BoseHubbard'}
\end{equation}
This Hamiltonian couples collective number states $\left| \overrightarrow{N}%
\right\rangle =\left| \ldots ,N_{-1},N_{0},N_{1},\ldots
\right\rangle $ where two neighboring wells $l$ and $l+1$ have
exchanged one particle. The
coupling between states with $N_{l}=p$, $N_{l+1}=q$ and $N_{l}=p-1$, $%
N_{l+1}=q+1$ is resonant for a harmonic $n$ when
\begin{equation}
n\omega +\kappa \left( q-p+1\right) =0,  \label{MultiResonance1}
\end{equation}
which is analogous to the resonance condition (\ref{rescond}). At
small tunneling rate we can exclude non-resonant coupling terms
while assuming that their effective coupling rate is negligible.
When the uniform force also contains a constant term, so that
$\varepsilon (t)=\varepsilon _{0}+\varepsilon _{1}\cos \omega t$,
we have to add a term $\varepsilon _{0}t $ to $\theta (t)$, and
the resonance condition is modified into
\begin{equation}
n\omega +\varepsilon _{0}+\kappa \left( q-p+1\right) =0.
\label{MultiResonance3}
\end{equation}
When $p-q=1$, this condition is independent of $\kappa $, and a
resonant oscillation can occur between states with
$N_{l}-N_{l+1}=\pm 1$.

Another interesting case is a Mott insulator state, with the same
number of particles $N_{0}$ in each well. Such a state has been
predicted in \cite {Jaksch} and has been recently experimentally
realized in (\cite{Greiner}), where one (two) atoms have been put
in a single lattice site. So, $\left| \Psi \left( t=0\right)
\right\rangle =\left| \ldots ,N_{0},N_{0},N_{0},\ldots
,\right\rangle .$ This state is directly coupled to the collective
Fock state which arises if a boson escapes to a
neighbouring well, so it has $N_{0}+1$ atoms in one lattice site, and $%
N_{0}-1$ in the neighboring one. Then the resonant condition is
$n\omega +\varepsilon _{0}+\kappa =0$.

Just as in the case of two wells, resonances coincide when $\omega
$ is of the same order as $\kappa $. When $\omega =\kappa $, there
is always a harmonic that resonantly couples neighboring wells. In
the absence of the constant term $\varepsilon _{0}$, the resonance
condition takes the universal form $n+q-p+1=0$. So, if in the Mott
insulator phase the number fluctuations are suppressed between
wells, we obtain their increase at resonances.

\section{Periodic modulation of coupling}

In this Section we consider the effects of a periodic modulation
of the coupling coefficient $\delta (t)$ between the wells. As a
simple model, we assume that $\delta $ contains a harmonic
component, so that
\begin{equation}
\delta \left( t\right) =\delta _{0}+\delta _{1}\cos \omega t.
\label{short0}
\end{equation}
In order that the even state $\left| g\right\rangle =(\left|
1\right\rangle +\left| 2\right\rangle )/\sqrt{2}$ is the ground
state, we keep $\delta (t)$
positive at all times, and we choose $\delta _{1}$ to be smaller than $%
\delta _{0}$. Hence we assume that $\delta _{0}\geq \delta
_{1}\geq 0$.
Moreover, we take the energy of the two wells to be equal, so that $%
\varepsilon =0$. The Hamiltonian in the form of (\ref{rHN}) with
the coupling coefficient (\ref{short0}) can be easily implemented
in practice. It describes a BEC in a two-well configuration with a
periodically modulated barrier height. Precise calculations of the
coupling coefficient are given in \cite{Salmond}.

Since in the Hamiltonian (\ref{rHN}) the term proportional to
$\widehat{J}_{x} $ is periodically modulated, we expect that the
basis of states $\left| \mu
\right\rangle _{x}$, which are eigenstates of the operator $\widehat{J}_{x}$%
, is the natural basis to describe the evolution. Then it is
convenient to
describe the Hamiltonian in terms of the operators $\widehat{J}_{x}$ and $%
\widehat{J}_{x}^{\pm }$, which are defined in Eq. (\ref{Jxplus}).
By using the identities
$\widehat{J}_{z}=-(\widehat{J}_{x}^{+}+\widehat{J}_{x}^{-})/2$
and $\widehat{J}_{x}^{+}\widehat{J}_{x}^{-}+\widehat{J}_{x}^{-}\widehat{J}%
_{x}^{+}=2[J(J+1)-\widehat{J}_{x}^{2}]$, we rewrite the
$N$-particle Hamiltonian (\ref{rHN}) in the form
\begin{equation}
\widehat{H}_{N}=-\hbar \delta \left( t\right)
\widehat{J}_{x}+\frac{\hbar
\kappa }{2}\left( J(J+1)-\widehat{J}_{x}^{2}\right) +%
{\displaystyle {\hbar \kappa  \over 4}}%
\left( \widehat{J}_{x}^{+2}+\widehat{J}_{x}^{-2}\right) .
\label{HNmjux}
\end{equation}
This expression demonstrates that a state $\left| \mu
\right\rangle _{x}$ is coupled only to its next nearest neighbors
$\left| \mu \pm 2\right\rangle _{x}$. The coupling strength is
measured by the matrix element
\begin{equation}
\Omega _{\mu }^{x}=\frac{\kappa }{4}\text{ }_{x}\left\langle \mu
+2\right|
\widehat{J}_{x}^{+2}\left| \mu \right\rangle _{x}=\frac{\kappa }{4}\sqrt{%
\left( J+\mu +1\right) \left( J+\mu +2\right) \left( J-\mu
-1\right) \left( J-\mu \right) },  \label{Couplingx}
\end{equation}
which depends on the interparticle interaction coefficient $\kappa
$ and the particle number $N=2J$.

In order to get a closer insight to the role of periodic
modulation and its resonances, we again eliminate the diagonal
part of the Hamiltonian, now with respect to the basis of states
$\left| \mu \right\rangle _{x}$. The time-dependent state is
expressed
\begin{equation}
\left| \Psi \left( t\right) \right\rangle =\widehat{S}(t)\left|
\Psi ^{\prime \prime }\left( t\right) \right\rangle ,
\label{intpicturex}
\end{equation}
with
\begin{equation}
\widehat{S}(t)=\exp \left[ i\eta \left( t\right) \widehat{J}_{x}-\frac{1}{2}%
i\kappa t\left( J(J+1)-\widehat{J}_{x}^{2}\right) \right] ,
\label{St}
\end{equation}
and $\eta \left( t\right) =\int_{0}^{t}dt^{^{\prime }}\delta
\left( t^{^{\prime }}\right) $ is the integrated coupling
coefficient. In order to obtain the Schr\"{o}dinger equation for
the transformed state $\left| \Psi ^{\prime \prime }\left(
t\right) \right\rangle $, we need the transformation
property of the off-diagonal operators $\widehat{J}_{x}^{+2}$ and $\widehat{J%
}_{x}^{-2}$. The transformed state $\left| \Psi ^{\prime \prime
}(t)\right\rangle $ obeys the Schr\"{o}dinger equation with the
effective Hamiltonian
\[
\widehat{H}^{\prime \prime }(t)=%
{\displaystyle {\hbar \kappa  \over 4}}%
\widehat{S}^{\dagger }(t)\left( \widehat{J}_{x}^{+2}+\widehat{J}%
_{x}^{-2}\right) \widehat{S}(t).
\]
In analogy to Eq. (\ref{LinearTransform}), we now apply the
general transformation rule
\begin{equation}
g(\widehat{J}_{x})\widehat{J}_{x}^{+2}=\widehat{J}_{x}^{+2}g(\widehat{J}%
_{x}+2),  \label{QuadTransform}
\end{equation}
for an arbitrary analytical function $g$ of $\widehat{J}_{x}$.
After making a Fourier expansion we obtain for
$\widehat{H}^{\prime \prime }(t)$ the explicit expression
\begin{equation}
\widehat{H}^{\prime \prime }(t)=\frac{\hbar \kappa
}{4}\sum_{n=-\infty }^{\infty \text{ }}J_{n}\left(
{\displaystyle {2\delta _{1} \over \omega }}%
\right) \left[ \widehat{J}_{x}^{+\text{ }2}e^{-it\left( \delta
_{0}+2\kappa \left( \widehat{J}_{x}+1\right) +n\omega \right)
}+\text{H.c.}\right] \label{Hx}
\end{equation}

The form of the Hamiltonian $\widehat{H}^{\prime \prime }(t)$
resembles the
Hamiltonian $\widehat{H}^{\prime }(t)$, as specified in Eq. (\ref{Hresonant}%
). In the present case, the basis states are the states $\left|
\mu \right\rangle _{x}$, which are now coupled by the square of
the corresponding ladder operator $\widehat{J}_{x}^{\pm \text{
}2}$. As in (\ref {Hresonant}), the coupling term is a series of
harmonics with equidistant frequencies $n\omega $, with an
amplitude proportional to the Bessel function of the corresponding
order.

In the high-frequency limit, when the modulation frequency $\omega
$ is
large compared with the diagonal frequency splittings of the Hamiltonian (%
\ref{HNmjux}), the effect of the static term proportional to
$J_{0}(2\delta _{1}/\omega )$ in Eq. (\ref{Hx}) will be dominant,
and the Hamiltonian will be effectively constant. Just as in
precious sections, the physical reason is that a rapidly modulated
field, which has a negligible average pulse area, also has a
negligible influence.

On the other hand, Eq. (\ref{Hx}) immediately shows that the
coupling between the state $\left| \mu \right\rangle _{x}$ and
$\left| \mu +2\right\rangle _{x}$ of the $n$th harmonic is
resonant when
\begin{equation}
n\omega +2\kappa (\mu +1)+\delta _{0}=0.  \label{resonancesX}
\end{equation}
The other coupling terms are negligible when the coupling strength
is small compared with the oscillation frequency, which leads to
the weak-coupling criterion
\begin{equation}
\left| \frac{\Omega _{\mu }^{x}}{n\omega +2\kappa \left( \mu
+1\right) +\delta _{0}}J_{n}\left(
{\displaystyle {2\delta _{1} \over \omega }}%
\right) \right| \ll 1.  \label{CriteriaX}
\end{equation}

\begin{figure}[!ht]
\centerline{\psfig{figure=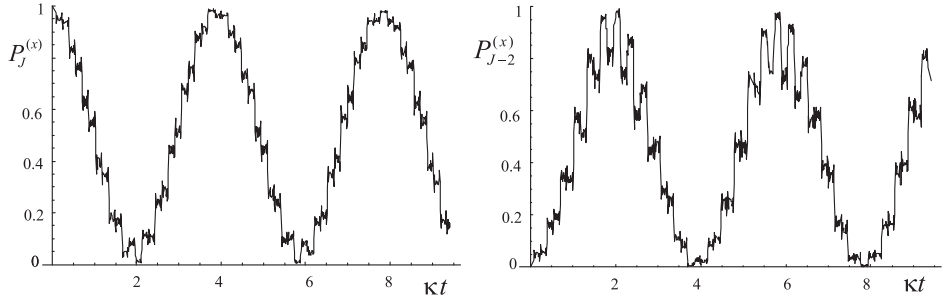,height=4cm}} \caption{Time
dependence of the populations $P_{\mu }^{\left( x\right) }$ of the
state $\left| \mu \right\rangle _{x}$ with $\mu =J$ and $\mu
=J-2$. The other parameters are taken as $\delta _{1}/\kappa =14$,
$\delta _{0}/\kappa =16$,  $\omega /\kappa =20$, $N=14$.}
\label{resf4}
\end{figure}

So, if the initial state $\left| \Psi \left( 0\right) \right\rangle =$ $%
\left| J\right\rangle _{x}$ is resonantly coupled to $\left|
J-2\right\rangle _{x}$, while further couplings of this alter
state are negligible, we have an effective two-level system. This
is demonstrated on the figure \ref{resf4}, where besides the
resonant oscillations, one obtains non-resonant escape of
population to the rest of manifold. Upon decreasing the coupling
between the state $\left| \mu \right\rangle _{x}$, only the
resonant states are involved and they exhibit clear Rabi
oscillations. This shows how resonances lead to an escape of
population from the initial state to the other states in the
manifold of states $\left| \mu \right\rangle _{x}$. The similarity
with the response to the periodic modulation in the form of a
periodically modulated energy difference between the wells. Recall
that the state $\left| J\right\rangle _{x}$ is the state in which
all particles are in the even state $\left| g\right\rangle
=(\left| 1\right\rangle +\left| 2\right\rangle )/\sqrt{2}$, which
is the one-particle ground state. In the state $\left|
J-2\right\rangle _{x}$, two particle have been transferred to the
odd excited state.

\section{Conclusions}

A BEC trapped in a two-well potential can be expected to be very
sensitive to the frequency of any applied periodic perturbation.
We test this idea by periodically modulating the asymmetry or the
barrier height of such a configuration. Compared with the
analogous situation of a single atom trapped in a light field with
a periodic modulation, the many-particle nature of the BEC gives
rise to some new effects. For both types of modulation, two-state
resonances may be observed, where a single atom out of the BEC
oscillates between the wells. It is also possible to enter a
regime of parameters where more than two states are resonantly
coupled, with more than one particle oscillating between the
wells. Using such resonances, one can manipulate the average
number of particles in the wells by varying the relevant
parameters, such as  the magnitude and the modulation frequency of
the energy difference. This effect can be considered also as a
means to resonantly enhance the tunnelling rate between wells. We
generalize the basic ideas developed for two-wells to a multiwell
system, such as a BEC in an optical lattice. Whereas the periodic
modulation of the energy difference is related to coupling between
number states in the two wells, the periodic modulation of the
height of the barrier leads to coupling between number states in
superposition states of the two wells with specific values of the
relative phase.

\acknowledgments
This work is part of the research program of the ``Stichting voor
Fundamenteel Onderzoek der Materie'' (FOM).

\end{document}